\documentclass[a4paper,11pt]{article}

\usepackage[margin=1.14in]{geometry} 
\usepackage{tabularx}
\usepackage{multirow}
\usepackage{hyperref}
\usepackage{booktabs}
\usepackage{color}
\usepackage{arydshln}
\usepackage{amsmath}
\usepackage{amsthm}
\usepackage{amssymb}
\usepackage{graphicx}
\usepackage{pdfpages}
\usepackage{float}
\usepackage{bm}
\usepackage{comment}
\usepackage{threeparttable}
\makeatletter
\makeatletter \renewcommand{\@dotsep}{10000} \makeatother
\usepackage{wrapfig}
\usepackage[utf8]{inputenc}
\usepackage{lmodern}
\usepackage{hyperref}
\usepackage{booktabs}  
\usepackage{forest}
\usepackage{geometry}
\usepackage{xcolor}
\usepackage{alltt}
\definecolor{desccolor}{gray}{0.45}
\usepackage{listings}
\forestset{
  dir tree/.style={
    for tree={
      font=\ttfamily\small,
      grow'=0,
      child anchor=west,
      parent anchor=south west,
      anchor=west,
      calign=first,
      edge path={
        \noexpand\path [draw, \forestoption{edge}]
        (!u.south west) +(7.5pt,0) |- node[fill,inner sep=1.25pt]{} (.child anchor)\forestoption{edge label};
      },
      before typesetting nodes={
        if n=1{insert before={[,phantom]}}{}
      },
      fit=band,
      before computing xy={l=15pt},
    },
  },
}

\definecolor{darkred}{rgb}{0.6, 0, 0}
\hypersetup{
	unicode,
	colorlinks,
	breaklinks,
	urlcolor=darkred, 
        linkcolor=black, 
        citecolor=darkred,
	pdfauthor={Author One, Author Two, Author Three},
	pdfproducer={LaTeX},
	pdfcreator={pdflatex}
}
\usepackage[T1]{fontenc} 
\usepackage{amsmath}

\newcommand{\be}{\begin{eqnarray}}
\newcommand{\ee}{\end{eqnarray}}
\def\be{\begin{equation}}
\def\ee{\end{equation}}
\def\bea{\begin{eqnarray}}
\def\eea{\end{eqnarray}}

\newcommand{\gsim}{\;\raisebox{-0.9ex}{$\textstyle\stackrel{\textstyle >}{\sim}$}\;}
\newcommand{\lsim}{\;\raisebox{-0.9ex}{$\textstyle\stackrel{\textstyle<}{\sim}$}\;}
\def\lsim{\raise0.3ex\hbox{$\;<$\kern-0.75em\raise-1.1ex\hbox{$\sim\;$}}}
\def\gsim{\raise0.3ex\hbox{$\;>$\kern-0.75em\raise-1.1ex\hbox{$\sim\;$}}}

\usepackage{graphics}

\usepackage{epsfig}
\usepackage{slashed}
\usepackage[utf8]{inputenc}
\usepackage{changepage}
\usepackage{multirow}
\usepackage{pstricks}
\usepackage{dcolumn}
\newcommand{\warnmark}{%
  \tikz[baseline=0.1ex, scale=0.18]{
    \draw[orange, line width=1.2pt, rounded corners=0.8pt]
      (0,0) -- (1,1.8) -- (2,0) -- cycle;
    \node[orange, font=\bfseries] at (1,0.70) {!};
  }%
}
\newcommand{\checkmarkgreen}{%
  \tikz[baseline=0.35ex, scale=0.18]{
    \draw[green!55!black, line width=1.2pt, rounded corners=1.2pt]
      (0,0) rectangle (2,2);
    \draw[green!55!black, line width=1.5pt, line cap=round, line join=round]
      (0.45,1.05) -- (0.85,0.60) -- (1.55,1.40);
  }%
}

\usepackage[sort&compress,numbers,square]{natbib}

\theoremstyle{plain}

\theoremstyle{definition}

\usepackage{graphicx, color}

\title{
Articulating Assumptions in AI-Generated Scientific Analyses through Task Decomposition
}

\vspace{8mm}\author{
\Large{
Ahmed Hammad$^{a}$\thanks{
email: hammad@kias.re.kr
}
\ and\
Mihoko Nojiri$^{b,c,d}$\thanks{
email: nojiri@post.kek.jp\\
\hspace*{2.em}Package:
\href{https://github.com/AHamamd150/LLM_For_scientific_Codes}{GitHub}}}}

\date{ 
{}
$^a$ Center of AI and Natural science, KIAS, Seoul 02455, Korea.\\
$^b$ Theory Center, IPNS, KEK,  1-1 Oho, Tsukuba, Ibaraki 305-0801, Japan.\\
$^c$ The Graduate University of Advanced Studies (Sokendai), 1-1 Oho, Tsukuba, Japan.\\
$^d$ Kavli IPMU (WPI), University of Tokyo, 5-1-5 Kashiwanoha, Kashiwa, Chiba 277-8583, Japan.
}
\begin{document}

\maketitle

\begin{abstract}

Scientific results produced by LLM-generated analysis code must be understandable and reproducible. However, uncertainty can arise at different stages of the process, both in the original natural language specification and in the generated implementation. As a result, even executable code may not provide a clear understanding of which quantities are being computed or which assumptions determine the final results.
To address this challenge, we introduce quantity-grounded semantic differencing, a multi-agent framework for analyzing and comparing scientific programs generated by LLMs. The framework assigns code generation, execution, tracing, and validation to separate agents, allowing it to reconstruct how key output quantities are produced and to identify differences between the intended analysis and the implemented code. We also introduce a module that inspects ambiguities in the initial user instruction and suggests alternative rewrites before code generation.
Its modular design is intended to facilitate application to different scientific domains by replacing domain-specific resources while preserving the same workflow. We validate the framework on representative collider physics analyses.  In the studied HEP cases, the modular task decomposition provides explicit intermediate artifacts for inspection, and the complete workflow was operated with substantially smaller models than in the previous implementation.

\end{abstract}

\newpage
\noindent\rule{\textwidth}{1pt}
\tableofcontents
\noindent\rule{\textwidth}{0.2pt}
\maketitle \flushbottom
\vspace{4mm}

\section{Introduction}
Large language models (LLMs) have recently achieved remarkable progress in generating executable code from natural language descriptions 
\cite{zan2023large,Jiang2024CodeGenerationSurvey,Ni2024L2CEval}. This capability is beginning to transform scientific computing by assisting not only in the development of new software but also in the modification of existing code and the construction of complex analysis pipelines \cite{Hou2024LLMSoftwareEngineering}. As a result, researchers no longer need to implement every computational step manually. Instead, they can describe computational tasks in natural language and integrate the generated code into a broader scientific workflow \cite{Umre:2026CSLLM}.

However, natural-language-based code generation introduces two distinct sources of uncertainty. The first arises from the request itself. Unlike programming languages, natural language does not impose a strict specification of computational procedures and often leaves room for interpretation in essential aspects of an analysis, including the target set of objects, exclusion criteria, aggregation units, and derivation rules \cite{zan2023large}. The second arises from the variability and fallibility of the stochastic generation process. Even for identical or closely related requests, a language model may produce different outputs, some of which may be inappropriate, incomplete, or semantically incorrect \cite{chen2021evaluatinglargelanguagemodels}. Consequently, the executability of generated code does not ensure that users understand what the output represents, nor does it make explicit the assumptions and choices that underlie the result.

This issue is particularly consequential in scientific computing, where numerical outputs must be accompanied by an account of what was computed, on which subset of the data, under what selection or exclusion criteria, and by which formulae or procedures the reported quantities were derived. Ensuring the validity and reproducibility of LLM-generated scientific analyses, therefore, requires exposing the quantities defined by the generated code and the computational paths through which final outputs are constructed.

The urgency of this challenge is particularly acute in high energy physics, where machine learning has become deeply integrated into core analysis workflows. Transformer based architectures have achieved strong  performance in jet classification and tagging tasks \cite{Qu:2022mxj,Bhimji:2025isp,Esmail:2025kii,Hammad:2024cae,Hammad:2023sbd}. Beyond individual taggers, recent work has explored the use of large language models and agentic systems directly within HEP analysis pipelines, including automated event selection, analysis card generation, and end-to-end workflow orchestration \cite{Heneka:2025fpe,Bakshi:2025fgx,Song:2025odk,Connor:2025zcy,Birk:2025fbs,Gendreau-Distler:2025fsj,Menzo:2025cim,Rafique:2025xeo,Zoccheddu:2026cjr,Hill:2026naa,Knipfer:2026kng,Atif:2026eju,Badea:2026klb,Li:2026krn,Mallampalli:2026hrl,Qiu:2026iby,Moreno:2026mqk,Agrawal:2026lvg,Menzo:2026qrl,Jat:2026yxv,Shih:2026jfe,Krzmanc:2026fdw,Saad:2026pan,Faroughy:2026dkj}. The growing use of AI-generated components in physics analyses makes the question addressed in this paper increasingly important. When analysis code is generated from a natural language description, it is not enough to verify that the code simply runs. It must also be possible to confirm that the generated code correctly implements the intended physics, that the object definitions and derived observables match the analysis specification, and that any assumptions introduced during code generation are explicit and can be inspected.

At the same time, general purpose and commercially hosted LLMs are not specifically optimized for scientific reasoning or individual physics domains, while their reliance on external services can limit local control and introduce recurring usage costs. This motivates the development of smaller, locally deployable models that can be adapted to scientific tasks through domain knowledge, physical constraints, and structured scientific information. In this work, we explore this direction using multiple cooperating small language model agents, with the aim of adapting agentic workflows to high-energy physics analysis rather than introducing the general concept of multi-agent LLM systems itself.

We propose a framework for inspecting computational code generated by LLMs from a user description of the desired output. Since such descriptions may leave important computational details unspecified, the framework is designed not only to generate executable code, but also to trace implementation choices that arise when the natural language request is translated into code. By making these choices explicit, the framework helps users understand what the generated code computes and how its outputs depend on assumptions introduced during translation.
To achieve this goal, we introduce a set of modules that mediate interactions among the user specification, the LLM, and the generated code. The \texttt{helper\_selector} module identifies predefined helper functions relevant to the requested task and injects them into the generation context, thereby constraining and guiding the code generation process. The \texttt{tracer} module analyzes the generated code and reconstructs an implemented specification, making explicit the code-level definitions of selected collections, event selections, composite quantities, histograms, and outputs. The \texttt{critique} module then compares this reconstructed specification with the original user request and highlights potential mismatches, hidden assumptions, or implementation choices that affect the results. 
Complementing these components, we introduce an ambiguity clarification module, referred to as the \texttt{Golden Axe Oracle} (\texttt{oracle}), which generates targeted clarification questions prior to code generation.

The framework is designed to be domain-adaptable: the generic workflow is kept independent of domain-specific resources, which are supplied through a fixed interface. In this study, we evaluate the framework by adapting the domain assets of \textsc{CoLLM} \cite{esmail2026collm}, originally designed for high energy physics analysis workflows. Whereas the original \textsc{CoLLM} framework was primarily developed and tested with LLMs at the 70B scale, our results indicate that the task decomposition can be used with substantially smaller models. The decomposed workflow was successfully operated with 14B-scale models in the present study. 

The remainder of this paper is organized as follows. Section 2 describes the proposed framework. Section 3 presents the experimental setup and evaluation protocol. Section 4 reports the main results and discusses recurring error patterns. Section 5 concludes the paper. A further refinement of the workflow and a detailed assessment of the frozen implementation are presented in the Appendix.

\section{Multi-Agent Framework}
We develop a multi-agent framework for generating, executing, and inspecting scientific analysis code. The framework transforms a natural language analysis specification into an executable implementation through a sequence of specialized agents responsible for helper selection, code generation, execution-based repair, tracing, and critique. By separating these responsibilities among distinct agents, the system produces intermediate representations that can be inspected, validated, and archived. This design reduces reliance on a single model output and improves the detection of implementation errors, ambiguous assumptions, and conceptual inconsistencies. These intermediate outputs can also provide feedback for refining domain-specific prompts and policies.

Domain-specific behavior is encapsulated within interchangeable packages that define generation prompts, helper libraries, input--output conventions, and agent policies. The orchestration logic remains unchanged when the system is adapted to a new scientific application. In the current implementation, domain packages are provided for particle physics analyses and lightweight testing scenarios, while the architecture is designed to be extensible to other domains.
A distinguishing feature of the framework is its combination of execution feedback with post-generation review. Generated code is executed in a runtime environment and may be iteratively repaired when failures occur. Successful implementations are subsequently analyzed by dedicated tracing and critique agents that evaluate the correspondence between the original specification and the final executable program. The resulting workflow integrates code generation, execution, repair, tracing, and automated review within a unified pipeline.

\subsection{Methodology}

\begin{figure}[ht]
    \centering
    \includegraphics[width=0.6\linewidth]{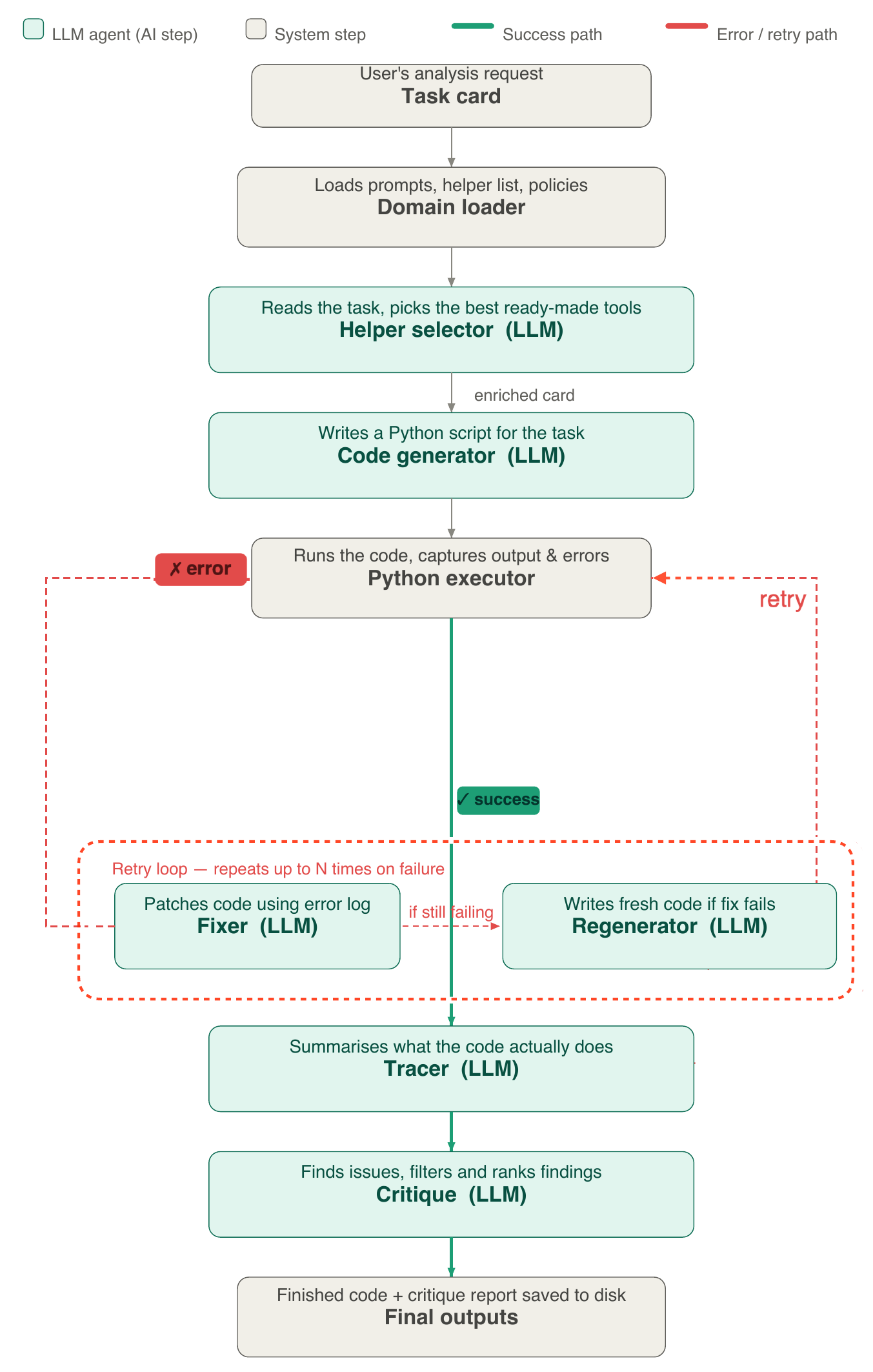}
    \caption{Schematic architecture of the multi-agent workflow.}
    \label{fig:network}
\end{figure}

The workflow consists of six LLM stages, as illustrated schematically in Figure~\ref{fig:network}. Each stage receives a well-defined input representation and produces structured intermediate artifacts that are passed to subsequent components. This design allows each step of the code generation process to be inspected independently and provides a clear separation between task interpretation and validation.

\paragraph{Domain initialization}

The workflow begins with initializing a domain package, which defines the scientific context in which the analysis is performed. The domain package separates domain-specific knowledge from the general orchestration logic and provides the information required by downstream agents through predefined formats. In this implementation, a domain package consists of five main components: a \texttt{generation profile}, a \texttt{helper registry}, a \texttt{helper selection policy}, a \texttt{critique context}, and an \texttt{input--output contract}.

The generation profile provides instructions for the code generation agent, including domain conventions, reusable helper functions, and analysis utilities. The helper registry provides structured metadata for the available helper functions, including their names, purposes, keywords, aliases, and dependencies. This registry is particularly important for smaller LLMs, since it allows the helper selection step to guide code generation without requiring the model to infer all available domain utilities from context alone. The helper selection policy and critique context define domain-specific guidance for the helper selection and critique agents, respectively. The input--output contract specifies the expected structure of inputs and outputs for the generated analysis code. This modular structure allows the same workflow to be applied to different scientific applications by replacing the domain package while preserving the underlying pipeline. In this way, domain-specific information is introduced only to the required points, without modifying the overall architecture.

\paragraph{Helper selection}

After initialization, the workflow receives an analysis specification describing the desired task
through an itemized \texttt{task card}. A helper selection agent (\texttt{helper selector}) analyzes this \texttt{task card} together with the available helper registry and identifies the utilities most relevant to the requested analysis. The original \texttt{task card} and the selected helper names are combined to form the {enriched} task card. A domain helper policy can also be inserted after the helper selector system prompt.  See Section 3 for a description of how the domain and system tasks are defined. 

The purpose of this stage is to recommend the selected helper functions as the candidates likely relevant to the requested task. 
The broader domain instruction and reusable utilities are supplied separately as the \texttt{generation profile}. 
This design guides the code generation agent toward plausible domain utilities and reduces unnecessary regeneration of helper functions, while leaving the final implementation strategy to the generator.
The \texttt{helper registry} may describe either helper functions whose source code is inserted into the generation context or helpers that are provided by an external library and imported by the generated program. In the latter case, the registry metadata must include not only the helper name and purpose, but also its expected inputs, outputs, dependencies, and import path, so that the code generation agent can use the helper without seeing its full implementation. 

\paragraph{Code generation}
The enriched \texttt{task card} is subsequently passed to the code generation agent, which produces a Python analysis script using the domain-specific \texttt{generation profile} and the selected helper candidates. The generation profile must specify how the LLM-generated code should read external data and write output. 
The generated code represents the first complete implementation of the requested analysis and serves as the input for the subsequent execution and review stages.
Since complex scientific analyses may exceed the output limits of current LLMs, the framework includes an automatic continuation mechanism. If the generated program is incomplete, an additional generation step is performed. 
The resulting program is then evaluated through direct execution.

\paragraph{Execution and iterative repair}

The generated program is executed as an independent Python subprocess to check whether it runs in the target environment. It runs with the permissions of the local execution environment and is not intended to provide a security sandbox; no additional filesystem, network, credential, or resource isolation is imposed. Failures produce runtime diagnostics that are used for correction.
When execution fails, the generated program, along with the runtime diagnostics, is passed to a dedicated \texttt{fixer} agent. The \texttt{fixer} attempts to produce a revised version of the implementation that resolves the observed failure, and the corrected program is executed again. If repeated repair attempts are unsuccessful, the workflow invokes a \texttt{regeneration} agent\footnote{The temperature of the regeneration agent is set to 0.3 by default. However, the temperature parameter of all other agents is set to zero.}, which produces a new implementation based on the enriched task card. 

This separation between \texttt{fixer} and \texttt{regeneration} allows the framework to address different classes of failures. Local programming errors can often be handled through targeted revision of the generated code, whereas more fundamental failures in the generating approach may require a new generation attempt. The execution--repair cycle continues until an executable implementation is obtained or a predefined retry limit is reached.

\paragraph{Implementation tracing}
Once generated code is available, it can be passed to a \texttt{tracer} agent together with the original \texttt{task card}. 
 The \texttt{tracer} constructs an as-implemented specification by extracting the code fragments relevant to each item in the \texttt{task card} and describing the corresponding code definitions.
For each requested quantity or selection in the \texttt{task card}, the \texttt{tracer} follows the relevant variable definitions and dependency chains back to their code-level origins, such as object collections, selection cuts, helper function calls, and derived quantities. 
We refer to this comparison between task-card quantities and their reconstructed code-level expressions, source spans, dependency chains, collection definitions, and composite structures as quantity-grounded semantic differencing.
The resulting structured representation makes the implemented program inspectable and reduces the burden on a downstream \texttt{critique} agent, especially when using smaller LLMs that may struggle to evaluate long source code directly. Unlike the other modules, the \texttt{tracer} does not require domain assets and is designed to be domain-independent.

The first step of the analysis is definition extraction. The \texttt{tracer} system prompt contains instructions not to stop at intermediate quantities under any circumstances.  The next stage is the definition check for the \texttt{task card} items.  Additional trace rules instruct the model to show the identified definition as a short chain "A computed as B $\rightarrow$ C $\rightarrow$ D".  The definition summary includes the identified chains and selected, filtered, sorted, or otherwise redefined collections in a fixed format. 

There are many failure modes in this tracer stage.  The system prompt is extensively validated with LHCO domain examples to veto failure modes, ensuring it works for LLMs with O(10B) parameters. The resulting tracer system prompt is rather long. We recommend that users not modify or reduce the system prompt without validation. We show the typical output in Section 4. 

\paragraph{Post-generation critique}

The final stage of the workflow is performed by a \texttt{critique} agent, which evaluates the consistency between the \texttt{task card} and the implemented program. The \texttt{critique} agent receives the original \texttt{task card}, the generated program, and the as-implemented specification produced by the \texttt{tracer}. The tracer output is used as the primary representation of the implemented workflow, while the source code is retained as supporting evidence when needed.
Using these inputs, the \texttt{critique} agent identifies potential issues, including missing requirements, discrepancies between the requested and implemented workflows, hidden assumptions or implementation choices that may affect the reported outputs. The resulting output is a structured findings report intended for human inspection.  The default configuration runs the critique stage once at zero temperature. 

\paragraph{Golden Axe Oracle}

The \texttt{Golden Axe Oracle} module  (\texttt{oracle}) detects ambiguities in the user input that may affect code design.\footnote{
The name refers to Aesop's ``Golden Axe'' fable. In the story, the goddess asks the woodcutter whether he had lost a golden axe or a silver axe, even though he had actually dropped an ordinary axe.}
Part of the instability in code generation and in helper selection is traced to ambiguities in the user prompt. 
The code generation process can handle such ambiguities gracefully, often producing executable code; however, the model's interpretation may remain implicit.
The \texttt{oracle} explains why the selected phrase is ambiguous and encourages users to resolve it. 
The module takes the \texttt{task card}, the \texttt{helper registry}, and a \texttt{domain oracle context} as input. A domain-independent instruction asks the module to inspect the \texttt{task card}  to identify at most three high-risk phrases that may affect the generated implementation. On the other hand, \texttt{domain oracle context} describes known ambiguity patterns.  For each selected phrase, the module explains why a particular task-card item is ambiguous, together with two possible interpretations and rewrites.  
We also provide \texttt{oraclet} (Oracle Templated), a templated variant of the \texttt{oracle}. Unlike the free-form \texttt{oracle}, \texttt{oraclet} asks the model to report ambiguities using a fixed set of ambiguity categories and includes more domain-dependent instructions. This variant is intended to make ambiguity inspection more stable, especially when using very small LLMs.

\vspace{4mm}
Throughout the workflow, all intermediate artifacts are stored in a run directory. These artifacts include selected helpers, intermediate outputs in the code generation processes, generated source code, execution outputs, error logs,  tracer and critique, and oracle outputs. Maintaining this complete record helps to improve the prompts. For example, the discovered common failure patterns can be generalized to improve system prompts and domain contexts. This provenance-oriented design supports auditability and may facilitate reproducibility and provide a transparent account of each stage of the workflow. Consequently, the framework supports not only automated generation of an analysis program but also systematic inspection and review of the resulting scientific workflow. 

In summary, the workflow provides a general framework for transforming high-level analysis specifications into executable scientific programs while preserving the intermediate artifacts needed for inspection. Unlike approaches that rely primarily on a single large proprietary model or an interactive chat-based coding process, the framework separates the generic orchestration logic from domain-specific resources and decomposes the task into explicit, auditable stages. By combining specialized agents for code design guided by helper selection, execution-based feedback, semantic reconstruction, post-generation review, and \texttt{task card} ambiguity review, the framework is designed to improve the transparency and reliability of LLM-generated analyses without making full autonomy the central objective. Moreover, the \texttt{domain critique and oracle contexts} allow freedom to customize/update the workflow for smaller LLMs. 

In the following section, we demonstrate the application of this framework to a particle collider analysis scenario in which the domain-specific components are instantiated for a high energy physics use case.

\section{A Case Study of Domain-Specific Tuning}

The generic orchestration layer is designed to be independent of a particular scientific domain.
 This independence is implemented through a domain asset interface. The workflow does not import domain-specific prompts or helper definitions directly; instead, it loads a \texttt{DomainAssets} object from a domain package specified by an import path such as \texttt{lhco\_domain}. In the current implementation, this object provides a generation profile, a helper registry, a helper selection policy, a critique context, and an input--output contract. The generation profile and helper registry are required fields, while the policy, critique context, and input--output contract provide optional guidance.

In this section, we introduce a complete set of domain assets for particle collider analyses for the LHC Olympics
(LHCO) event format \cite{Thaler:2006LHCO}. The LHCO event format is a lightweight text representation of collider events in terms of reconstructed objects such as jets, leptons, photons, taus, and missing transverse momentum.  The LHCO domain assets provide domain vocabulary and recurring ambiguity patterns for the \texttt{critique} and \texttt{Golden Axe Oracle} analyses. 
The LHCO domain assets are derived from the domain resources used in the original \textsc{CoLLM} framework and reorganized to fit the domain asset interface used in the present workbench. 

\subsection{Code generation profile}

The generation profile defines the main code generation instructions for LHCO collider analyses. It provides a substantial domain-specific prompt rather than a minimal set of hints. The profile specifies that the generated program should assume LHCO input, include the provided LHCO parser, and use the allowed numerical and plotting libraries to produce runnable Python analysis code. It also defines collider analysis conventions for object types, leading and subleading objects, missing transverse momentum, four-momentum construction, invariant mass, transverse mass, angular separation, cutflow bookkeeping, histogram production, and output formatting.

In addition to these general conventions, the profile includes a library of reusable helper functions for parsing, object selection, kinematic reconstruction, composite object construction, multiplicity counting, charge handling, and common resonance candidate definitions.  
These helpers provide the baseline domain infrastructure available to the code generation agent. The system prompt instructs the LLM to preserve its names, signatures, return values, and internal logic when copied into the generated code. The model assumes that \texttt{[ADDITIONAL\_HELPERS]} guidance is provided with the \texttt{task card}. The LHCO generation profile instructs the model to treat these selected helpers as the { primary helper guidance} for the current analysis.  
Note that the embedded helper library is now substantially simpler than the original \textsc{CoLLM} implementation because the helper dependencies on the base helpers are made explicit. This becomes possible with a reliable, well-organized selection of helpers using the \texttt{helper selector}.   

The profile is therefore domain specific in content, but generic in its role within the workflow: it supplies the code generation agent with the reusable conventions, parser, and helper infrastructure required by the target application, while allowing the helper selector and user specification to determine which resources are emphasized for a particular analysis.
\subsection{Helper registry and helper selection policy}

The helper registry provides structured metadata for reusable helper functions that the code generator can use. Each entry describes a helper function by its name, category, short summary, keywords, aliases, and dependencies. The registry serves as the searchable index used by the helper selection agent. Since the complete helper library may contain many functions that are irrelevant to a given analysis, the helper selector recommends task-specific helper candidates, while the generation profile and domain package provide the broader code generation context.

In the present LHCO implementation, many reusable helpers are still embedded in the generation profile or copied into the generated code. This design keeps the generated program self-contained and is useful for small-scale experiments. As the helper collection grows, however, the same interface can be adapted to a library-based design in which helpers are imported rather than inserted verbatim.

For the LHCO domain, the helper selection policy provides short domain-specific guidance for selecting helper functions from the registry. The policy instructs the model to pay attention to explicit cuts, vetoes, reconstruction steps, and requested observables in the task specification. On the other hand, the system side of the prompt emphasizes exact helper names and explicit relevance to the requested analysis, and discourages selecting helpers only because of superficial keyword overlap. It also instructs the model not to invent, rename, shorten, or alias helper names.
Some of these constraints overlap with the domain-independent helper selection instructions used by the workbench. This repetition is intentional: for small LLMs, restating key constraints in the domain-specific policy helps keep the selector focused on the explicit requirements of the task and reduces unsupported helper selection.

\subsection{Critique context}

The generic critique prompt remains domain-independent and provides the rules on handling \texttt{task card} and \texttt{tracer} output. The LHCO critique context supplies the collider analysis vocabulary and semantic expectations used by the post-generation critique stage. The context identifies the domain as LHCO collider analysis and lists basic semantic objects that may appear in such analyses, including reconstructed particle collections, selected object collections, missing transverse momentum, cutflow counters, histograms, reconstructed composite systems, and output quantities.

In addition, the LHCO context defines domain rules for composite quantities and for choices between selected collections and broader event level collections. Those are some cases that frequently appear as ambiguities or semantic mismatches in the generated programs. The context also provides the six finding types used by the critique agent: (A) composite violation, (B) definition mismatch, (C) missing implementation, (D) specification ambiguity, (E) extra behavior, and (F) generator-induced bias. 

\subsection{Oracle context}

The oracle context provides domain-specific guidance for ambiguity clarification by capturing ambiguity patterns known to affect code generation. Rather than detecting all possible linguistic ambiguities, the ambiguity clarification module focuses only on relevant domain expressions that may lead to multiple executable implementations. The overall interaction pattern remains domain-independent. When an ambiguity is identified, the oracle context presents the possible interpretations of the user's request and asks the user to select the intended meaning. The ambiguity patterns and clarification instructions are defined by domain experts to ensure accurate code generation for the target software package and workflow.

For the LHCO domain, the oracle context includes the following recurring sources of ambiguity:
\begin{itemize}
\item whether an object refers to a selected collection or to a raw event level collection;
\item whether a veto or multiplicity requirement is applied to selected objects or to all event level objects;
\item whether a phrase refers to a reconstructed composite object or only to a scalar quantity;
\item whether validation histograms are filled after the full event selection or as soon as the observable becomes available;
\item whether leading and subleading objects are defined before or after selection cuts;
\item whether the same physical object is reused consistently across cuts, validation plots, and outputs;
\item whether multiple helpers in the registry can compute similar quantities on different object collections, while the user input does not specify which collection should be used.
\end{itemize}

In addition, we provide two variants of \texttt{oracle}. One of them is \texttt{oracle-lite}, which does not contain LHCO specific examples and serves as a less domain-dependent template. The other is \texttt{oraclet}, a templated variant of \texttt{oracle} that constrains both the categories of ambiguity and the output types. It is intended to make ambiguity reports more structured and stable, especially for smaller LLMs. In the following validation, we mostly present the results using the \texttt{oracle}.  

\section{Workflow Validation}

In this section, we evaluate the different stages of the proposed workflow, from helper selection and code generation to semantic validation and ambiguity detection. The objective is not only to determine whether executable code can be produced but also to assess whether the generated implementation faithfully represents the user's analysis specification.

To this end, we construct a benchmark consisting of five representative collider physics analyses, summarized in Table~\ref{tab:user_inputs_summary}. Each benchmark is provided to the framework through a structured \texttt{task card} containing three components: the event selection requirements, a set of validation observables, and the desired output structure. The selected analyses cover a variety of reconstruction patterns, including diboson production, semileptonic top quark reconstruction, diphoton resonances, electroweak gauge boson production, and vector boson fusion Higgs topologies.

\begin{table}[!h]
\centering
\small
\begin{tabular}{p{0.03\linewidth}p{0.25\linewidth}p{0.34\linewidth}p{0.30\linewidth}}
\hline
Case & Physics process & Main selection cuts & Special definitions / structured instructions \\
\hline

1 &
\(pp \to W^+W^-\), 
\(W^+ \to \ell^+\nu\), 
\(W^- \to jj\)
&
Exactly one lepton; at least two jets; MET requirement; 
\(m_T(\ell,\mathrm{MET})\) window for the leptonic \(W\); 
dijet mass window for the hadronic \(W\); 
\(\Delta R(j_1,j_2)<3.0\); \(b\)-jet veto.
&
The hadronic \(W\) candidate is represented by the two leading jets.
The validation/output observables include the dijet system and
\(\Delta\phi\) between the lepton and the two jet system.
\\

\hline

2 &
\(pp \to t\bar{t}\), 
\(t \to bW^+\), 
\(W^+ \to \ell^+\nu\), 
\(\bar{t}\to \bar{b}W^-\), 
\(W^- \to jj\)
&
Exactly one lepton; at least four jets; at least two \(b\)-tagged jets;
MET requirement; \(m_T(\ell,\mathrm{MET})\) window; 
hadronic \(W\)-mass window.
&
The hadronic \(W\) candidate is built from the two leading non-\(b\)-tagged jets.
The hadronic top candidate is built from the leading \(b\)-jet and the
hadronic \(W\) candidate.
\\

\hline

3 &
\(pp \to H\), 
\(H \to \gamma\gamma\)
&
At least two photons; leading/subleading photon \(p_T\) requirements;
diphoton invariant mass window; 
\(p_T^{\gamma_1}/m_{\gamma\gamma}>0.35\);
\(p_T^{\gamma_2}/m_{\gamma\gamma}>0.25\).
&
The leading and subleading photons are defined after photon selection.
The diphoton system is explicitly built from these two photons.
The same diphoton system is used to define both \(m_{\gamma\gamma}\)
and \(p_T^{\gamma\gamma}\).
\\

\hline

4 &
\(pp \to WZ\), \(W\to \ell\nu\), \(Z\to \ell^+\ell^-\)
&
Exactly two selected electrons and one selected muon; opposite charge
electron pair; \(Z\)-mass window; MET requirement; W-candidate muon
\(p_T\) requirement; \(m_T(\mu,\mathrm{MET})>30~\mathrm{GeV}\).
&
The \(Z\) candidate, \(z_{\rm obj}\), is built from the two selected electrons
using summed four momentum.
The selected muon is explicitly defined as the \(W\)-candidate lepton.
The validation plots include the three lepton system.
\\

\hline

5 &
\(pp \to Hjj\), 
\(H \to \tau^+\tau^-\)
&
At least two taus; at least two jets; opposite-sign tau pair;
\(m_{jj}>400~\mathrm{GeV}\); 
\(|\Delta\eta(j_1,j_2)|>2.5\);
ditau mass window; \(b\)-jet veto.
&
The leading/subleading jets (taus) are defined from selected jets (taus) and used to build the dijet (ditau) system.
Composite observables include \(\Delta\phi(\tau\tau,jj)\) and the centrality of the ditau system relative to the two jets.
\\
\hline
\end{tabular}
\caption{
Summary of the five user inputs.  Standard object level requirements such as
basic \(p_T\) and \(\eta\) cuts are omitted unless they define a special object
or reconstruction step.  The table emphasizes the main event selection logic
and the structured instructions for reconstructed or composite objects.
}
\label{tab:user_inputs_summary}
\end{table}

\subsection{Preparation of benchmark task cards}

Preliminary studies revealed that the quality and reproducibility of generated code depend strongly on the structure of the analysis specification itself \cite{Akli:2026,Khojah:2024,Paleyes:2025}. We therefore refined the benchmark \texttt{task card}  to reduce semantic ambiguities and to provide a more explicit description of the intended reconstruction procedure. The most important modification concerns the treatment of composite physics objects. Whenever an observable or selection criterion depends on a reconstructed candidate, the \texttt{task card} explicitly instructs the model to construct that candidate as a reusable object obtained by summing the four momenta of its constituents. For example, in the VBF $(H\rightarrow\tau^+\tau^-)$ benchmark (Case~5), the card specifies that the ditau system must be constructed from the selected leading and subleading taus, while the dijet system must be constructed from the selected leading and subleading jets. Subsequent cuts and validation observables are then defined in terms of these reconstructed systems rather than through repeated direct calculations. A similar strategy is adopted throughout the benchmark suite. In the $H\rightarrow\gamma\gamma$ analysis (Case~3), for example, the diphoton system is explicitly defined before any invariant mass or transverse momentum requirements are applied. The \texttt{task card} therefore specifies not only which observables should be evaluated, but also how intermediate reconstructed objects should be represented and reused during code generation.

These modifications improved generation stability, particularly for smaller models such as Qwen2.5-14B and Qwen3-14B. Explicit structural instruction defining composite objects improved the consistency of generated implementations across repeated generations. This observation highlights a central feature of the framework: the \texttt{task card} functions not only as a physics specification but also as an explicit communication layer between the user and the language model. By expressing physics assumptions directly within the \texttt{task card} rather than implicitly through the system prompt, the framework improves transparency and facilitates the identification of discrepancies between the intended and implemented analyses. A further refinement of the workflow and task specifications, together with a detailed evaluation of the frozen implementation, is presented in the Appendix.

\subsection{Stability of helper selection}

The first stage of the workflow consists of helper selection. This stage is motivated by a practical limitation we observed in a previous CoLLM study \cite{esmail2026collm}: when the code generation agent is given only the initial task card, the \texttt{code generation}–\texttt{fixer} loop rarely produces a usable implementation for a model as small as 14B parameters. These failures are varied, including incorrect helper usage, omission of required helper calls, and ad hoc re-implementation of helper functionality instead of using the provided utilities. Explicitly separating helper identification from code generation reduced the observed instability in our tests, allowing the workflow to operate with 14B-parameter models, whereas the previous CoLLM implementation was primarily demonstrated with models of approximately 70B parameters. In the current workflow, the \texttt{helper selector} agent takes the \texttt{task card} and a helper registry as input and identifies the subset of predefined utilities most relevant to the requested analysis.
Since subsequent code generation relies heavily on the selected helper list, the stability of this stage is an important indicator of the robustness of the overall workflow.

We evaluate the stability of helper selection using Qwen2.5-14B and Llama3-70B. For each benchmark analysis, the selector is executed independently five times.
Table~\ref{tab:helper-selector-stability} summarizes the number of helpers selected consistently across all runs, together with the run-to-run variation in the total number of selected helpers.
\begin{table}[!h]
\centering
\begin{tabular}{lcccccc}
\hline
Model & Case & Common & Non common & Min/run & Max/run & Jaccard value  \\
\hline
Qwen2.5-14B  & 1 & 25 & 2 & 25 & 27 & 0.956\\
 & 2 & 22 & 7 & 24 & 27 & 0.872\\
 & 3 & 12 & 3 & 13 & 14 & 0.919\\
 & 4 & 14 & 3 & 15 & 16 & 0.929 \\
& 5 & 16 & 1 & 16 & 17 & 0.976 \\
 \hline
Llama3-70B & 1 & 23 & 2 & 24 & 24 & 0.952 \\
 & 2 & 22 & 1 & 22 & 23 & 0.983\\
 & 3 & 13 & 0 & 13 & 13 & 1.000\\
 & 4 & 14 & 0 & 14 & 14 &1.000\\
  & 5 & 18 & 2 & 18 & 20 & 0.950 \\
\hline
\end{tabular}
\caption{Stability of helper selection over five selector runs. ``Common'' denotes helpers selected in all five runs, while ``non-common'' denotes helpers selected in only a subset of runs. ``Min/run'' and ``Max/run'' show the minimum and maximum number of helpers selected in a single run for each model and analysis case.}
\label{tab:helper-selector-stability}
\end{table}
Table ~\ref{tab:helper-selector-stability} is intended to quantify the reproducibility of helper selection across repeated runs, rather than to compare the overall performance of the two LLMs. Although both models select a similar total number of helpers, Llama3-70B generally exhibits smaller run-to-run variation than Qwen2.5-14B. Importantly, this stability should not be interpreted as a measure of correctness, since different helper sets may lead to equally valid implementations. The average Jaccard similarity is also reported. Overall, Llama3-70B exhibits substantially greater stability than Qwen2.5-14B. In most cases, only one or two helpers differ across repeated Llama-70B runs, whereas Qwen2.5-14B shows noticeably larger fluctuations. Nevertheless, stability alone should not be interpreted as a measure of correctness. Multiple helper sets may be compatible with the same \texttt{task card}; a stable selection may simply reflect model-specific bias rather than an objectively optimal choice. The LLMs are trained using publicly available code; therefore, the helper choice may be influenced by standard HEP analyses. This is why the program validation stage is necessary. 

\subsection{Validation of generated code}

We validate code generation on the five benchmark cases summarized in Table~1. Since the qualitative patterns are similar across the benchmark, we present Case~5 as a representative detailed example. This case provides a useful example because it combines composite object reconstruction with validation observables that involve relationships among reconstructed systems. For each helper selector model, we used the most frequently generated helper set as the representative helper configuration. This corresponds to 16 helpers for Qwen2.5-14B and 20 helpers for Llama3-70B, excluding the mandatory LHCO parser. Table~\ref{tab:case5-cleaned-codegen-eval-vertical} summarizes the semantic evaluation of four representative generated implementations
using the legacy Case~5 task input.

The implementation using Qwen2.5-14B for both helper selection and code generation is classified as a partial success. The main event selection requirements are implemented, but one validation observable is inconsistent with the \texttt{task card}. In particular, for $\Delta\phi(\tau\tau,jj)$, the generated code computes the azimuthal separation between the leading tau and the leading jet, rather than the separation between the ditau system and the dijet system.

By contrast, the other three implementations correctly reproduce $\Delta\phi(\tau\tau,jj)$ and the ditau centrality. In particular, the Llama3-70B code generator correctly implements these composite observables with either helper selector. This suggests that the quality of the generated implementation depends not only on helper selection but also on the code generation model.

Minor discrepancies remain even in otherwise successful implementations. For example, with the Llama3-70B helper selector, the 14B-generated code restricts the $b$-jet veto using the $p_T$ and $\eta$ requirements, whereas the 70B-generated code applies the veto to the raw jet collection. In addition, some histograms in the 14B-generated code are filled before the final selection, while others are filled after it. Some generated implementations also omit the specified histogram ranges.

The legacy task input does not explicitly specify which jet collection to use for the $b$-jet veto, nor does it define the precise point at which validation histograms should be filled. These implementation differences illustrate ambiguities in the task specification and motivate the \texttt{tracer} and \texttt{critique} stages discussed later in this study.

\begin{table}[!h]
\centering
\begin{tabular}{lll}
\hline
\texttt{helper selector}/\texttt{codegen} & Item & Evaluation \\
\hline

Llama3-70B  / Llama3-70B 
                          & Core selection & pass \\
                          & Composite $\Delta\phi(\tau\tau,jj)$ & pass \\
                          & Ditau centrality & pass \\
                          & Minor issues & b-jets veto, plot ranges  \\
                          & Overall & pass \\
\hline
Llama3-70B  / Qwen2.5-14B 
                         & Core selection & pass \\
                         & Composite $\Delta\phi(\tau\tau,jj)$ & pass \\
                         & Ditau centrality & pass \\
                         & Minor issues & plot fill timing is partial \\
                         & Overall & pass \\
\hline
Qwen2.5-14B  / Llama3-70B 
                         & Core selection & pass \\
                         & Composite $\Delta\phi(\tau\tau,jj)$ &  pass \\
                         & Ditau centrality & pass \\
                         & Minor issues &  plot ranges, cutflow labels \\
                         & Overall & pass  \\
\hline
Qwen2.5-14B  / Qwen2.5-14B 
                        & Core selection & mostly pass \\
                        & Composite $\Delta\phi(\tau\tau,jj)$ & fail: uses leading $\tau$ and leading jet \\
                        & Ditau centrality & pass \\
                        & Minor issues & 
                        cut order differs; incomplete cutflow \\
                        & Overall & partial success \\
\hline
\end{tabular}
\caption{Manual semantic evaluation of four Case~5 code-generation outputs using the legacy task input (\texttt{user\_input\_5\_legacy.txt}).
}
\label{tab:case5-cleaned-codegen-eval-vertical}
\end{table}

\subsection{Ambiguity Detection}

Natural language analysis specifications often admit multiple valid interpretations, leading to different implementations. If these ambiguities are resolved implicitly by the language model, the generated analysis code may no longer reflect the user's intended workflow. To avoid this failure mode, \texttt{ oracle} performs an explicit ambiguity detection stage before any code is generated. As described already, rather than committing to a single interpretation, the model identifies potentially ambiguous statements and asks the user to resolve them. For every detected ambiguity, the model reports the original phrase, explains why the ambiguity may affect the generated analysis, asks a clarification question, and proposes two possible interpretations. These are referred to as the \texttt{Golden Axe} and the \texttt{Silver Axe}. The \texttt{Golden Axe} corresponds to the recommended or conventional interpretation whenever such a preference exists, whereas the \texttt{Silver Axe} provides an alternative but equally plausible reading. This protocol makes the semantic choices explicit and allows the user to resolve ambiguities before code generation begins.

We first evaluate a free-form prompting strategy, referred to as \texttt{oracle-lite}, in which the model is instructed to discover ambiguities without predefined categories. Table~\ref{tab:legacy_oracle_lite_recount}  summarizes the ambiguities identified by different language models over ten independent runs. We classified each report as useful, partially useful, or off-target. The classification was first assisted by ChatGPT and subsequently reviewed by the authors. The exact ChatGPT model version and the interactive prompt history used for this legacy curation were not frozen and cannot be fully reconstructed. Since each run allows at most three ambiguity reports, the maximum achievable score is thirty valid detections.\footnote{The legacy run data and the prompt are available in the GitHub release with a summary CSV and additional Python scripts. }

The legacy free-form oracle-lite experiment on the original Case~2 task card produced 172 ambiguity reports across 60 runs, out of a maximum of 180. To distinguish the recognition of implementation-sensitive choices from the ability to formulate actionable clarification questions, we classified each report as useful, partially useful, or off-target (Table~\ref{tab:legacy_oracle_lite_recount}).
The classification gives 38 useful, 79 partially useful, and 55 off-target reports, with eight output slots left unfilled. For Qwen2.5-32B, only two reports were classified as useful while 17 were partially useful; for Qwen2.5-72B, the corresponding counts were 17 and six. This contrast suggests that recognizing a potentially consequential choice and expressing it as a well-posed clarification question are distinct challenges.

\begin{table}[ht]
\centering
\setlength{\tabcolsep}{6pt}
\begin{tabular}{lrrrrr}
\toprule
Model & Useful & Partially useful & Off-target & No report & Reported \\
\midrule
Qwen2.5-7B & 1 & 12 & 16 & 1 & 29 \\
Qwen2.5-14B & 2 & 13 & 15 & 0 & 30 \\
Qwen2.5-32B & 2 & 17 & 4 & 7 & 23 \\
Qwen2.5-72B & 17 & 6 & 7 & 0 & 30 \\
\midrule
Llama3.1-8B & 3 & 18 & 9 & 0 & 30 \\
Llama3.3-70B & 13 & 13 & 4 & 0 & 30 \\
\midrule
Total & 38 & 79 & 55 & 8 & 172 \\
\bottomrule
\end{tabular}
\caption{Classification of the legacy free-form oracle-lite outputs on the original Case~2 task card. Each model was run ten times, with at most three ambiguity reports per run. Useful reports pose a well-formed clarification about an unresolved implementation choice; partially useful reports identify a relevant implementation distinction but have an incomplete or imperfect clarification or ask about a choice already specified by the task card. Off-target reports do not provide a useful implementation distinction. No report counts unfilled output slots, rather than incorrect responses. }
\label{tab:legacy_oracle_lite_recount}
\end{table}

\begin{table}[t]
\centering
\setlength{\tabcolsep}{7pt}
\begin{tabular}{lrrrr}
\toprule
Reported issue category & Useful & Partially useful & Off-target & Total \\
\midrule
$W$ collection scope & 13 & 12 & 0 & 25 \\
$b$-jet collection scope & 7 & 10 & 1 & 18 \\
$W$ candidate reuse & 1 & 14 & 4 & 19 \\
Top candidate consistency & 0 & 6 & 2 & 8 \\
Lepton definition & 14 & 12 & 3 & 29 \\
Selection ordering & 0 & 7 & 35 & 42 \\
Output timing & 1 & 0 & 0 & 1 \\
MET consistency & 0 & 2 & 0 & 2 \\
Other & 2 & 16 & 10 & 28 \\
\midrule
Total & 38 & 79 & 55 & 172 \\
\bottomrule
\end{tabular}
\caption{ Issue-category breakdown of the 172 legacy oracle-lite reports, pooled across all six models. These are report-level counts; categories are assigned for analysis and should not be interpreted as distinct task-card ambiguities. The Other category contains reports not covered by the original eight categories.}
\label{tab:legacy_oracle_lite_categories}
\end{table}

The legacy results show model-dependent differences.  Useful claims appear in Qwen2.5-72B and Llama3.3-70B, identifying 30 valid ambiguities. The smaller Qwen 2.5-7B, Qwen2.5-14B and Llama3.1-8B produced 40 off-target reports.  We treat them as descriptive because the comparison was not performed under a fully controlled inference budget. 

Table \ref{tab:legacy_oracle_lite_categories} shows the issue category of all model runs, where useful issues appear in the W collection scope and lepton definition.  The ambiguities discovered by the models can be grouped into several recurring categories.

\begin{enumerate}
\item \textbf{Collection scope.} It may be unclear whether an object is selected from the original collection or from a subset that has already
passed previous selection requirements. Typical examples include the definition of the leading or subleading $b$-jet and the jet collection used to reconstruct the hadronic $W$ candidate.

\begin{quote}
Are the two leading non-\texttt{b}-tagged jets selected from the full jet collection, or only from the jets that satisfy the previous \texttt{pT}/\texttt{eta} requirements?
\end{quote}

\item \textbf{Candidate consistency.} Reconstructed objects may either be reused throughout the analysis or rebuilt independently at later stages. This ambiguity commonly appears when constructing hadronic $W$ and top candidates.

\begin{quote}
Should the hadronic $W$ candidate constructed earlier be reused when building the hadronic top candidate?
\end{quote}

\item \textbf{Object definition.} Ambiguities may arise from the definition of composite object collections, such as whether the requirement of exactly one lepton applies to the combined electron--muon collection or independently to each flavour.

\begin{quote}
Do you require exactly one lepton from the combined electron and muon collections, or exactly one electron or exactly one muon?
\end{quote}

\item \textbf{Selection ordering.} The interpretation of a requirement may depend on whether it is applied before or after the previous kinematic selections, for example, when counting leptons after the individual \texttt{pT} and $\eta$ requirements have already been imposed.

\item \textbf{Output timing.} Validation histograms and derived quantities may either correspond to the final selected events or be filled as soon as the relevant quantities become available.

\item \textbf{Variable consistency.} The same reconstructed quantity should be used consistently throughout the analysis. A typical example is ensuring that the missing transverse momentum entering the event selection is identical to that used in the transverse mass calculation.
\end{enumerate}

Although the \texttt{oracle-lite} prompt provides considerable flexibility, it also places a substantial reasoning burden on smaller language models. These models frequently overlook important ambiguities or produce observations that are only weakly related to the user's analysis. To make ambiguity discovery more structured and reduce the reasoning burden on smaller models, we introduce domain \texttt{oracle} and \texttt{oraclet } (\texttt{Oracle Templated}), which replace unconstrained ambiguity discovery by adding domain-specific examples (\texttt{oracle}) or a fixed set of semantic categories (\texttt{oraclet}). In particular, \texttt{oraclet} explicitly specifies three categories to examine: collection scope, multi-part object, and output timing. 

\subsection{Tracer and Critique}

The tracer generally produced structured and readable outputs at the 14B scale, while the frozen evaluation also revealed recurring fidelity limitations. We present our representative example here; a systematic 50-run tracer audit is given in Appendix Section 6.4.  As an illustration, we consider the diphoton analysis of Case~3, whose \texttt{task card} explicitly defines the leading and subleading photons after photon selection and requires the diphoton system to be built from these two objects. An excerpt of the selection block of the \texttt{task card} reads:
\begin{lstlisting}[
  basicstyle=\ttfamily\small,
  breaklines=true,
  frame=single
]
[SELECTION_CUTS]
- Select photons with pT > 25 GeV and |eta| < 2.37.
- Exclude photons in the crack region 1.37 < |eta| < 1.52.
- Require at least 2 photons
- Define the leading photon and subleading photon after photon selection.
- Leading photon pT > 35 GeV
- Subleading photon pT > 25 GeV
- Build the diphoton system from the leading photon and the subleading photon.
....
....
\end{lstlisting}
 
Given this specification and the generated code, the tracer reconstructs the as-implemented specification item by item. For each requirement in the \texttt{task card}, it reports the corresponding code expression, the upstream chain of variable definitions from which the expression is derived, and the source line references that ground the reconstruction in concrete code evidence. The example tracer output generated by  Qwen3-14B for Case~3  reads as  follows:
\begin{lstlisting}[
  basicstyle=\ttfamily\small,escapechar=!,
  breaklines=true,
  frame=single
]
!\checkmarkgreen! Select photons with pT > 25 GeV and |eta| < 2.37.; computed as: [p for p in photons if p['pt'] > 25 and abs(p['eta']) < 2.37]; upstream chain: photons -> get_photons(event) [L252-L254]
!\checkmarkgreen! Exclude photons in the crack region 1.37 < |eta| < 1.52.; computed as: [p for p in photons if not (1.37 < abs(p['eta']) < 1.52)]; upstream chain: photons -> get_photons(event) [L252-L254]
!\checkmarkgreen!  Require at least 2 photons; computed as: len(photons) >= 2; upstream chain: photons -> get_photons(event) [L252-L254]
!\checkmarkgreen! Define the leading photon and subleading photon after photon selection; computed as: helper_get_leading(photons_sorted) and helper_get_subleading(photons_sorted); upstream chain: photons_sorted -> sort_by_pt(photons) [L264-L266]
!\checkmarkgreen!  Leading photon pT > 35 GeV; computed as: leading_photon['pt'] > 35; upstream chain: leading_photon -> helper_get_leading(photons_sorted) [L271-L274]
....
....
* Build the diphoton system from the leading photon and the subleading photon.; computed as: sum_objects(leading_photon, subleading_photon); upstream chain: leading_photon, subleading_photon -> helper_get_leading, helper_get_subleading [L281-L282]
...
...
* [CHECK CAREFULLY] DEFINITION SUMMARY
* photons
  - computed/defined as: [o for o in event['objects'] if o['type'] == 0]
  - upstream chain: event['objects']
  - collection definition: o['type'] == 0
  - structure: single-constituent
  ....
  ....
* diphoton
  - computed/defined as: sum_objects(leading_photon, subleading_photon)
  - upstream chain: leading_photon, subleading_photon -> helper_get_leading, helper_get_subleading
  - collection definition: sum_objects
  - structure: composite-derived
....
....
\end{lstlisting}
 
The tracer output consists of two parts. The first part maps each item in the \texttt{task card} to the code fragment that implements it, along with its upstream dependency chain and source-line references. This mapping makes it possible to verify, for example, that the kinematic photon selection is applied to the collection returned by the LHCO parser, and that the leading and subleading photons are defined from the sorted collection of selected photons rather than from the raw event level collection. The second part is a definition summary of each named object in the program, such as its defining expression, upstream chain, collection definition, and structural classification. In the example above, the \texttt{photons} collection is classified as a single constituent collection extracted directly from the event record, whereas the \texttt{diphoton} object is classified as a composite derived object built from the leading and subleading photons through the \texttt{sum\_objects} helper. This classification directly addresses the composite object ambiguities discussed in the previous subsections, since it makes explicit whether a named candidate is represented as a reusable combined object or exists only implicitly through repeated scalar computations.
 
The \texttt{critique} stage uses this reconstruction as its primary representation of the implemented workflow, while retaining the generated source code as supporting evidence. As a consequence, the quality of the critique output depends strongly on the quality of the tracer output. When the \texttt{tracer} correctly identifies the object collections and dependency chains, the \texttt{critique} agent can compare the as-implemented specification against the \texttt{task card} at the level of semantic quantities, and can attribute a finding, such as a definition mismatch or a composite violation, to a specific code expression and source-code location. Delegating the reconstruction of the implemented analysis to a dedicated tracing stage reduces the reasoning burden placed on the \texttt{critique}, which is particularly important for smaller models that struggle to evaluate long source programs directly. The structured, line-referenced format of the tracer output also allows the critique findings to be verified by human inspection, in line with the provenance-oriented design of the framework.

The critique output should be interpreted as a ranked inspection report rather than as an automatic pass/fail judgment. Each finding is assigned both a finding type and a severity marker. The finding type, such as composite violation or definition mismatch, describes the kind of semantic issue being considered. The severity marker indicates how strongly the issue should be treated in the final review. In the examples below, \color{orange}\warnmark{}\color{black} \ denotes a likely implementation mismatch, \color{red}?\ \color{black} denotes an ambiguity in the task specification, and '$\circledcirc$' denotes a low-severity finding retained for human inspection. Thus, a finding labeled as a composite violation may still be non-blocking if the implemented computation is semantically equivalent or if the issue only reflects a conservative critique rule.

In the following, we show representative examples generated by Qwen3-14B for Cases~3 and~5. For Case~3, the \texttt{critique} agent does not find major problems and summarizes the implementation with $\circledcirc$. For Case~5, the implementation contains collection-scope issues, which are marked by \color{orange}\warnmark{}\color{black}\ and \color{red}?\color{black}.\footnote{In the example below, SPEC refers to items in the task card.}

\noindent {\bf Case~3}

\begin{lstlisting}[
  basicstyle=\ttfamily\small,
  breaklines=true,escapechar=!,
  frame=single
]
[KEPT 1] keep: no obvious noise signal
!$\circledcirc$! (A) composite violation

SPEC:
- Build the diphoton system from the leading photon and the subleading photon.
- Define m_gammagamma as the invariant mass of this diphoton system.
- Use this same diphoton system to compute diphoton pT.

CODE:
- diphoton = sum_objects(leading_photon, subleading_photon)
- m_gammagamma = diphoton['mass'] if diphoton else 0.0
- px = leading_photon['pt'] * math.cos(leading_photon['phi']) + subleading_photon['pt'] * math.cos(subleading_photon['phi'])
- py = leading_photon['pt'] * math.sin(leading_photon['phi']) + subleading_photon['pt'] * math.sin(subleading_photon['phi'])
- diphoton_pt = math.sqrt(px**2 + py**2)

Reason:
The diphoton pT is computed directly from the transverse momenta of the individual photons rather than from the composite diphoton object. This is a composite violation because the SPEC requires the diphoton pT to be computed from the diphoton system, not from the individual photons.

Evidence priority:
- diphoton_pt is computed from px and py derived from leading_photon and subleading_photon, not from the diphoton object [L308-L312]
\end{lstlisting}

\noindent {\bf Case~5}\footnote{The Case~5 example is taken from a code-generation run that did not execute successfully and is used here to illustrate the semantic inspection performed by the tracer–critique workflow.}

\begin{lstlisting}[
  basicstyle=\ttfamily\small,escapechar=!,
  breaklines=true,
  frame=single
]
[KEPT 1] keep: no obvious noise signal
!\color{orange}\warnmark \color{orange}! (B) definition mismatch

SPEC:
- Build dijet system from the leading jet and subleading jet.

CODE:
- dijet_mass(event)

Reason:
The SPEC requires building a dijet system from the leading and subleading jets, but the code uses the helper function dijet_mass(event), which computes the invariant mass of the two leading jets. This does not explicitly construct the dijet system as a composite object, which may be required for downstream calculations.

Evidence priority:
- The dijet system is computed as dijet_mass(event) [L317]

...
...
[KEPT 5] keep: no obvious noise signal
!\color{red}?\color{red}! (D) specification ambiguity

SPEC:
- Veto events with b-tagged jets

CODE:
- n_bjets(event, pt_cut=30, eta_cut=4.7) == 0

Reason:
The SPEC does not specify whether the b-jet veto should apply to the selected jets (with pT > 30 GeV and |eta| < 4.7) or to the broader event-level jets. The code uses the selected jets, which is a plausible interpretation, but this choice is not explicitly constrained by the SPEC.

Evidence priority:
- The b-jet veto is computed as n_bjets(event, pt_cut=30, eta_cut=4.7) == 0 [L329-L331]

\end{lstlisting}

\section{Conclusion}

The reliability of LLM-generated scientific analyses depends not only on whether the generated code executes successfully, but also on whether it faithfully implements the physics intended by the researcher. These two criteria are not equivalent, and conflating them can introduce silent errors that propagate undetected through the scientific workflow. This work has addressed that gap by introducing a multi-agent framework in which code generation, execution, semantic reconstruction, and post-generation review are assigned to distinct agents operating through a structured pipeline. A central finding of this study is that decomposing the generation task into explicit, auditable stages provides additional opportunities to detect implementation failures that would otherwise remain invisible to standard execution-based validation. 

The \texttt{helper selector} selects the relevant helper functions to build the code,  making the helper-selection decision explicit before actual code generation starts.  The \texttt{tracer} and \texttt{critique} modules make it possible to evaluate whether the quantities computed by the generated program correspond to the quantities specified by the researcher, rather than merely whether the program terminates without error. This distinction is essential in collider physics analyses, where the physical interpretation of a result depends on precise definitions of object collections, composite observables, selection criteria, and cut flow ordering that may not be recoverable from the code alone.

A complementary finding concerns the role of the analysis specification itself. The results demonstrate that  explicit instruction of composite object construction within the \texttt{task card} was associated with reduced implementation variability in the studied cases across repeated generation attempts. This indicates that the \texttt{task card} functions as more than a physics description; it serves as a communication interface between the researcher and the language model, and its precision directly conditions the semantic fidelity of the generated implementation. The \texttt{Golden Axe Oracle} module (\texttt{oracle}) addresses the upstream source of this variability by identifying ambiguities in the user specification before code generation begins. By presenting alternative interpretations and requesting explicit clarification, the module displaces implicit resolution choices from the language model to the researcher, where they belong.

Together, these components establish a framework in which scientific code generation is treated not as a single inference step but as a structured process with multiple opportunities for inspection and correction. The intermediate artifacts produced at each stage, including the generation card, the as-implemented specification, and the critique report, constitute a transparent record of the assumptions that determine the final result. Such provenance supports auditability and may facilitate reproducibility in scientific workflows that must withstand independent scrutiny.

The framework operates with small LLMs of approximately 14B parameters in the sense that the workflow can proceed through code generation, execution, repair, tracing, and critique. In our experiments, this was achieved using the task-decomposed workflow introduced in this paper. A central technical contribution is the introduction of the \texttt{helper registry} and \texttt{helper selector}, which enable code generation with models as small as Qwen3-14B, whereas our previous implementation required models of approximately 70B parameters. This also makes local deployment feasible in principle and broadens the range of environments in which the workflow can be used.

The modular architecture of the framework, in which domain-specific resources are supplied through a fixed interface without modifying the orchestration logic, is designed to make the package accessible across scientific domains beyond high energy physics. The provided code-generation profile is specific to LHCO code generation, but  it provides some insights into creating a new domain profile. The helper selection, critique, and oracle system prompts are less dependent on domain knowledge, and they may be easier to adapt to other domains by removing domain-specific examples and adding appropriate domain information. 

The package is available on \texttt{GitHub}. Compared with the version used in the experiments, the released package includes adjustments to the user inputs and the helper registry so as to make the generated implementations more standardized. The version used in the experiments is also included as a legacy version.  Although the experiments reported in this paper were performed with a version prior to the domain-separation refactoring, the substantive functionality is the same. In addition, the present study is limited to relatively compact single-script analyses, and more complex workflows involving multiple files or long dependency chains may require additional decomposition. The current implementation also executes code locally and has not yet been validated on HPC systems. Extending the framework to HPC environments would require an execution backend for job submission, resource allocation, monitoring, and retrieval of logs and outputs, while leaving the tracing and critique stages unchanged.


\section*{Acknowledgments}
 This work is funded by grant number 22H05113, ``Foundation of Machine Learning Physics'', Grant in Aid for Transformative Research Areas and 22K03626, Grant in Aid for Scientific Research (C) and  JST, ARiSE, Japan Grant Number JPMJAR2633. A. Hammad is partially supported by the Science, Technology \& Innovation Funding Authority (STDF) under project ID 50806.
\section{Appendix: Current workbench performance}
\subsection{Frozen workflow}
The main text describes the workbench and reports results obtained with the legacy domain system prompts (version 1). The current frozen workflow (version 2) uses revised system prompts, and its performance is evaluated in this appendix. A major change is the removal of potentially confusing helper definitions inherited from CoLLM, with the aim of improving helper-selection stability.
At the same time, the task cards contain more operational statements than the previous CoLLM task cards. Here we describe the results of the frozen version of the package and describe the success rate of each step in detail. 

In Table \ref{tab:case_complexity}, we list the numbers of selection entries, operational entries, and validation plots in each card. 
To characterize the practical execution time of the current workflow, we performed ten independent runs for each of the five analysis cases, for a total of 50 runs. We measured end-to-end wall clock time from workflow invocation to completion of the final output, including helper selection, code generation, execution and repair when required, tracing, critique, the post-run Oracle review, API communication, and provider-side latency. Qwen3-14B is used throughout the generation. Code generation, execution and tracing were completed successfully for all 50 runs.  All 50 runs succeeded on the initial execution attempt, with no fixer or regeneration calls. In one run of Case~2, the critique and Oracle outputs were not produced because of an external HTTP 502 gateway error.\footnote{For the four representative legacy Case~5 outputs reported in Table 3, three succeeded on the initial execution attempt, while one Qwen2.5-14B/Qwen2.5-14B run required one fixer call before succeeding; no regeneration was required.}

\begin{table}[!ht]
\centering
\begin{tabular*}{0.7\textwidth}{@{\extracolsep{\fill}}lccc@{}}
\toprule
Case &
Selection entries &
Operational entries &
Validation plots \\
\midrule
1 & 10 & 1 & 10 \\
2 & 10 & 2 & 10 \\
3 & 12 & 4 & 9 \\
4 & 10 & 2 & 10 \\
5 & 13 & 4 & 10 \\
\bottomrule
\end{tabular*}
\caption{Complexity of the five benchmark task cards. Operational entries, included in the selection entries, correspond to object construction or definition steps.}
\label{tab:case_complexity}
\end{table}

The mean end-to-end latency over the 49 successfully completed runs was \(92.3 \pm 19.7\) s, where the uncertainty denotes the standard deviation across runs; the median was 86.6 s. The case-by-case mean latencies ranged from 73.6 s to 101.7 s, while individual successful runs ranged from 64.7 s to 161.8 s. Because these measurements include network communication and provider-side service latency, they should be interpreted as observed user-facing wall-clock times rather than intrinsic model inference times.
In addition to the five original task cards, we include Case~6 and Case~7, an improved version of Case~2 and Case~5. The number of selection entries and operational entries slightly increased for Case~6, as shown in Table \ref{tab:case_complexity_improved}, and changes are listed in Tables \ref{tab:case2_case6_spec} and \ref{tab:case5_case7_spec}. The collection scope of the b-jets and non-b-tag jets is more explicitly written in Case~6. For Case~7, we clarify the object definitions and their use in the subsequent steps, and we adjust the validation range of  \(m_{jj}\). These modifications improved generation stability in the studied cases. 

\begin{table}[t]
\centering
\begin{tabular*}{0.85\textwidth}{@{\extracolsep{\fill}}lccc@{}}
\toprule
Case & Selection entries & Operational entries & Validation plots \\
\midrule
6 & 12 & 3 & 10 \\
7 & 13 & 4 & 10 \\
\bottomrule
\end{tabular*}
\caption{Numbers of selection entries, operational entries among selection entries, and validation plots for Case~6 and Case~7.
Case~6 (Case~7) is an improved task card of Case~2 (Case~5), respectively.}
\label{tab:case_complexity_improved}
\end{table}

\begin{table}[!ht]
\centering
\small

\begin{tabular}{p{0.22\linewidth}p{0.34\linewidth}p{0.36\linewidth}}
\toprule
Item & Case~2 & Case~6 \\
\midrule

$b$-jet requirement
&
Require at least two $b$-tagged jets.
&
Among the selected jets, require at least two $b$-tagged jets.
\\

Closest $b$-jet
&
Used only in the requested $\Delta R$ plot; the relevant
$b$-jet collection is not explicitly defined.
&
Explicitly define the $b$-jet closest to the selected lepton
among the selected $b$-tagged jets.
\\

Hadronic $W$ candidate
&
Built from the two leading non-$b$-tagged jets.
&
Built from the two leading {selected} non-$b$-tagged jets.
\\

Hadronic top candidate
&
Built from the leading $b$-jet and the hadronic $W$ candidate.
&
Built from the leading {selected} $b$-tagged jet and the
hadronic $W$ candidate.
\\

Top mass range
&
$100$--$250$ GeV appears only as a validation plot range.
&
$100$--$250$ GeV is explicitly specified as a selection cut
and also used as the validation plot range.
\\

$\Delta R(\ell,b)$ plot
&
Leading lepton vs.\ closest $b$-jet.
&
Leading lepton vs.\ closest {selected} $b$-jet.
\\

\bottomrule
\end{tabular}
\caption{
Main specification changes from Case~2 to the clarified Case~6.
Case~6 makes the object provenance and reconstruction steps explicit,
while keeping the same underlying semileptonic top quark analysis as Case~2.
}
\label{tab:case2_case6_spec}
\end{table}

\begin{table}[!ht]
\centering

\begin{tabular*}{0.95\textwidth}{@{\extracolsep{\fill}}p{0.20\textwidth}p{0.34\textwidth}p{0.34\textwidth}@{}}
\toprule
Item & Case~5 & Case~7 \\
\midrule

$b$-jet veto
&
Veto events with $b$-tagged jets.
&
Veto events with $b$-tagged jets among the selected jets.
\\

$\Delta\phi(\tau\tau,jj)$
&
Plot the azimuthal separation between the ditau system
and the dijet system.
&
Plot the azimuthal separation explicitly between
\texttt{ditau\_obj} and \texttt{dijet\_obj}.
\\

Dijet mass validation range
&
$200 < m_{jj} < 2000~\mathrm{GeV}$.
&
$400 < m_{jj} < 2000~\mathrm{GeV}$,
consistent with the selection $m_{jj}>400~\mathrm{GeV}$.
\\

\bottomrule
\end{tabular*}
\caption{
Main specification changes from Case~5 to the clarified Case~7.
The underlying $Hjj$, $H\to\tau^+\tau^-$ analysis and the main
event selection requirements are unchanged; Case~7 clarifies the object provenance, the object scope used by the selected observables, and the $b$-jet veto.
}
\label{tab:case5_case7_spec}

\end{table}

\subsection{Helper selection stability under refined system prompts}

The helper selector defines the relevant helper set and adds it to the task card.  Table~\ref{tab:helper_stability_frozen}
summarizes the variation of the selected helper sets. After confusing helpers are removed, Cases~1, 3, and 5 select identical helper sets in all ten runs. Cases 2 and 4 showed limited variation, with mean pairwise Jaccard similarities of 0.949 and 0.970, respectively. 
In addition, we examined the number of events after the final selection in each generated analysis. For Cases~1, 3, 4, and 5, the terminal event count was identical in all ten runs. For Case~2, nine runs yielded 1205 events, while one run (Run 7) yielded 1777 events. For the nine runs, the requested plotting range of 100--250 GeV for the hadronic top mass was additionally interpreted as an event-selection cut, even though it is not stated in the task card. Run 7 did not apply this cut. Run 7 has a separate problem: the generated code computed the top mass but did not append it to the list used for the corresponding histogram, resulting in an empty plot. 

\begin{table}[!ht]
\centering
\begin{tabular}{crrrrrr}
\hline
Case & Runs & Common & Non-common & Min/run & Max/run & Mean Jaccard \\
\hline
1 & 10 & 22 & 0 & 22 & 22 & 1.000 \\
2 & 10 & 20 & 2 & 20 & 22 & 0.949 \\
3 & 10 & 14 & 0 & 14 & 14 & 1.000 \\
4 & 10 & 17 & 3 & 18 & 19 & 0.970 \\
5 & 10 & 16 & 0 & 16 & 16 & 1.000 \\
\hline
\end{tabular}
\caption{Helper selection stability in the frozen revision workflow.
The mean pairwise Jaccard similarity is computed over all unordered pairs
of the ten runs for each case. The metric quantifies run-to-run stability
only and does not measure helper relevance or correctness. Temperature is set to 0 for all runs.}
\label{tab:helper_stability_frozen}
\end{table}

\subsection {Implementation Fidelity  beyond selection stability}
Although the cutflow yields are numerically stable,  the agreement at the event counts does not by itself establish whether the generated analysis faithfully implements the intended semantics. We therefore examined Case~2, which exhibited the largest implementation-level variation, and compared ten generations obtained with the 14B and 32B models to ten 32B generations using the revised Case~6 task card.
Increasing the model size alone did not eliminate the recurring implementation issues. In particular, the $100$--$250$~GeV plotting range for the hadronic-top mass was interpreted as an additional event-selection requirement in 9/10 runs with the 14B model and in all 10/10 runs with the 32B model.  Similarly, the requested $\Delta R$ with respect to the closest $b$ jet was evaluated using the leading $b$ jet in 8/10 runs for both model sizes. In Case~6, these ambiguities were removed by explicitly defining the closest selected $b$ jet before its use and by promoting the hadronic-top mass window to an explicit selection requirement consistent with the corresponding validation plot. None of the three targeted implementation issues was observed in the ten Case~6 runs. The corresponding frequencies are summarized in Table~\ref{tab:case2_implementation}.

\begin{table}[ht]
\centering
\begin{tabular*}{0.95\textwidth}{@{\extracolsep{\fill}}lccc@{}}
\toprule
Implementation issue & Case~2 & Case~2 & Case~6 \\
& 14B & 32B & 32B \\
\midrule
Leading $b$ jet used instead of closest $b$ jet & 8 & 8 & 0 \\
Top-mass range inconsistent with task-card semantics & 9 & 10 & 0 \\
Top-mass histogram not populated & 1 & 3 & 0 \\
\midrule
Runs with $\geq 1$ listed issue & 10 & 10 & 0 \\
\bottomrule
\end{tabular*}
\caption{
Implementation issues in Case~2 (top-quark analysis), reported as the number of affected runs out of ten. Case~2 is the original task card, while Case~6 explicitly defines the closest selected $b$ jet and the hadronic-top mass selection.
}
\label{tab:case2_implementation}
\end{table}

\subsection{Tracer and critique evaluation}
After the helper selection and code generation, the tracer evaluates the generated code.  
We used GPT-5.6-Luna to pre-annotate tracer outputs at the level of individual task-card requirements. The screening covered requirement coverage, source correctness, dependency correctness, and hallucination. These annotations were not treated as ground truth; they were subsequently reviewed and adjudicated by human evaluators. We initially inspected one reference run from each case and an additional numerically distinct run (Case~2, Run~7) to define the expert adjudication criteria before extending the assessment of the full 50-run outputs. The six adjudication rules (E1 $\sim$ E6) are identified to specify the cases not counted as tracer errors, as follows: 

\begin{enumerate}
\item[\textbf{E1}] {\bf semantic equivalence}:
Semantically equivalent descriptions of the implemented computation are accepted; literal reproduction of the source code is not
required.

\item[\textbf{E2}] {\bf caller-side helper}: The caller-side helper call is sufficient. Recursive expansion of helpers called inside that helper is not required.

\item[\textbf{E3}] {\bf plotted quantity/provenance}: For plotted quantities, identifying the scientific quantity and its provenance is sufficient; citation of the plotting call or plotting configuration itself is not required.

\item[\textbf{E4}] {\bf tracing-granularity ambiguity}: The tracer is not penalized when the required source location or tracing granularity is not uniquely specified, including cutflow ordering, printing, and similar bookkeeping.

\item[\textbf{E5}] {\bf task-code mismatch}: A mismatch between the task card and the generated code is not itself a tracer error; the generated code is treated as the ground truth for what was actually implemented.

\item[\textbf{E6}] {\bf cut boundary semantics}: Selection requirements implemented through rejection conditions and \texttt{continue} are excluded from the primary fidelity assessment and evaluated separately for control-flow semantics. A trace is counted as
successful if it reports either the surrounding rejection-and-\texttt{continue} statement or the corresponding effective pass condition. Reporting the rejection condition alone, or only the computed quantity without the selection condition, is counted as a failure.
\end{enumerate}

Table \ref{tab:tracer_fidelity_14b} summarizes the numbers of hallucinations, dependency errors, source errors, and coverage misses after applying the five adjudication rules E1 to E5.  Because the partial success case is important, the evaluation of E6 is presented separately. The counts of the applied adjudication rules are  (E1, E2, E3, E4, E5)$= (23,2, 336, 136, 57)$. 
All Case~5 runs contain the same tracer hallucination: the tracer claims that explicit ditau and dijet objects are defined, even though the generated code does not define them. We reran the tracer--critique stage on the same generated analysis code using Qwen3-32B and found the hallucinated claims  were no longer introduced. The hallucination in Case~2 occurs in Run 7 and is associated with the treatment of the top mass list and the corresponding plot. The frozen Qwen3-14B tracer--critique run did not correctly identify this issue. 

\begin{table*}[ht]
\centering

\small
\begin{tabular}{lrrrrrrr}
\toprule
Case &
Runs &
Req.\ items &
Runs with  &
Hallucination &
Dependency  &
Source  &
Coverage  \\
&&& $\geq 1$ issue&&error &error&miss \\
\midrule
1 & 10 & 247 & 0  & 0  & 0  & 0 & 0  \\
2 & 10 & 238 & 4  & 8  & 1  & 0 & 10 \\
3 & 10 & 311 & 1  & 0  & 0  & 1 & 0  \\
4 & 10 & 238 & 0  & 0  & 0  & 0 & 0  \\
5 & 10 & 269 & 10 & 20 & 0 & 0 & 0  \\
\midrule
Total & 50 & 1303 & 15 & 28 & 1 & 1 & 10 \\
\bottomrule
\end{tabular}
\caption{
Tracer-fidelity assessment for the Qwen3-14B runs after applying the E1--E5 adjudication rules. The “Runs with $\geq 1$ tracer issue” column counts affected runs, whereas the remaining error columns count affected requirement-level items. Counts are reported at the task-card requirement level. Automated screening was used only as pre-annotation. 
}
\label{tab:tracer_fidelity_14b}
\end{table*}

Cut-boundary identification under E6 is also evaluated for the selected items listed in Table \ref{tab:e6_conditions}. The semantic completeness of the tracer is summarized in Table  \ref{tab:continue_semantics} and Table \ref{tab:e6_boundary_fidelity} for all 5 cases, namely a maximum of 70 items for a two-sided window and 20 items for a one-sided window. A trace is classified as {complete semantic capture} when it reports either the rejection condition together with the corresponding  {continue} control flow, or the equivalent effective pass condition. {Partial: condition only} denotes cases in which the rejection condition is reported without the associated control-flow semantics. {Partial: quantity only} denotes cases in which the relevant computed quantity is identified but the selection condition itself is not reported. The boundary recovery rate ( Pass+Partial)/Total is significantly low for Case~2. The corresponding recovery rate is 100\% for Case~6 with Qwen3-32B.

\begin{table}[!ht]
\centering
\begin{tabular*}{0.95\textwidth}{@{\extracolsep{\fill}}cll@{}}
\toprule
Case & Quantity / requirement & Evaluated boundary \\
\midrule
1 & Transverse mass of lepton + MET & $40 < m_{\mathrm{T}}(\ell,\mathrm{MET}) < 120~\mathrm{GeV}$ \\
1 & Invariant mass of the two leading jets & $60 < m_{jj} < 100~\mathrm{GeV}$ \\
\midrule
2 & Hadronic $W$ mass & $40 < m_{W}^{\mathrm{had}} < 120~\mathrm{GeV}$ \\
2 & Leading $b$-jet + hadronic-$W$ invariant mass & $100 < m_{t}^{\mathrm{had}} < 250~\mathrm{GeV}$ \\
\midrule
3 & Diphoton invariant mass & $105 < m_{\gamma\gamma} < 160~\mathrm{GeV}$ \\
\midrule
4 & Invariant mass of \texttt{z\_obj} & $76 < m(\texttt{z\_obj}) < 106~\mathrm{GeV}$ \\
\midrule
5 & Ditau invariant mass & $60 < m_{\tau\tau} < 150~\mathrm{GeV}$ \\
\midrule
6 & Hadronic $W$ mass & $40 < m_{W}^{\mathrm{had}} < 120~\mathrm{GeV}$ \\
6 & Leading $b$-jet + hadronic-$W$ invariant mass & $100 < m_{t}^{\mathrm{had}} < 250~\mathrm{GeV}$ \\
\midrule
1 & MET & $\mathrm{MET}>30~\mathrm{GeV}$ \\
2 & MET & $\mathrm{MET}>20~\mathrm{GeV}$ \\
\bottomrule
\end{tabular*}
\caption{Numerical boundaries used for the E6 evaluation. 
Case~6 uses the same physics boundaries as Case~2.}
\label{tab:e6_conditions}
\end{table}

\begin{table}[t]
\centering
\begin{tabular*}{0.9\textwidth}{@{\extracolsep{\fill}}lccccc@{}}
\toprule
Cut type & Items & Complete & Partial: & Partial: & N/A \\
& & semantic capture & condition only & quantity only & \\
\midrule
Two-sided window & 70 & 37 & 17 & 15 & 1 \\
Lower-bound only & 20 & 12 & 8 & 0 & 0 \\
\midrule
Total & 90 & 49 & 25 & 15 & 1 \\
\bottomrule
\end{tabular*}
\caption{Semantic completeness of tracer outputs for event-selection cuts treated under rule E6. These items are excluded from the primary tracer-fidelity errors in Table~\ref{tab:tracer_fidelity_14b}. N/A denotes boundaries not implemented in the generated code.}
\label{tab:continue_semantics}
\end{table}


\begin{table}[!ht]
\centering
\begin{tabular*}{0.9\textwidth}{@{\extracolsep{\fill}}llrrrrrr@{}}
\toprule
Model & Case & Pass & Partial & Fail & N/A & Evaluated & Recovery  \\
& & & & & & (excl. N/A) & rate \\
\midrule
14B & 1 & 14 & 0 & 6 & 0 & 20 & 70.0\% \\
14B & 2 & 2 & 8 & 9 & 1 & 19 & 52.6\% \\
14B & 3 & 10 & 0 & 0 & 0 & 10 & 100.0\% \\
14B & 4 & 1 & 9 & 0 & 0 & 10 & 100.0\% \\
14B & 5 & 10 & 0 & 0 & 0 & 10 & 100.0\% \\
\midrule
14B & Total & 37 & 17 & 15 & 1 & 69 & 78.3\% \\
\midrule
\midrule 
32B & 2 & 9 & 1 & 10 & 0 & 20 & 50.0\% \\
32B & 6 & 20 & 0 & 0 & 0 & 20 & 100.0\% \\
\bottomrule
\end{tabular*}
\caption{E6 selection-boundary fidelity. Pass denotes correct boundaries and selection semantics; Partial denotes correct boundaries with incomplete control-flow semantics; Fail denotes an incorrect or missing implemented boundary. N/A denotes boundaries not implemented in the generated code and excluded from the denominator. Recovery rate is (Pass+ Partial) /  Evaluated.}
\label{tab:e6_boundary_fidelity}
\end{table}

We evaluate critique detection for selected failure cases. We focus on Case~2 and Case~5, which contain failure modes due to the complexity. 
Table \ref{tab:case2_case6_improvement} shows generation error, tracer exposure, and critique detection for the quantities calculated after the selection cuts for Case~2 and Case~6. These instructions appear in the later section of the task cards; therefore, they provide a useful test of the handling of the context in the moderate-sized LLM.  
The Case~6 task card significantly reduced the evaluated generation errors when used with the 32B model. 
The situation is similar for Case~5 and Case~7 in Table \ref{tab:case5_case7_comparison}. 

\begin{table}[!ht]
\centering
\begin{tabular*}{0.98\textwidth}{@{\extracolsep{\fill}}llccp{0.20\textwidth}@{}}
\toprule
Issue & Setting & Generation  & Tracer & Critique  \\
& & error & exposure & detection\\
\midrule
Closest $b$-jet & Case~2 / 14B & 8/10 & 8/8 & 1/8 \\
& Case~6 / 14B & 0/10 & N/A & N/A \\
& Case~6 / 32B & 0/10 & N/A & N/A \\
\midrule
Muon $\lvert\eta\rvert$ threshold mismatch & Case~2 / 14B & 0/10 & N/A & N/A \\
& Case~6 / 14B & 10/10 & 10/10 & 9/10 \\
& Case~6 / 32B & 0/10 & N/A & N/A \\
\midrule
Unrequested top-mass cut & Case~2 / 14B & 9/10 & 0/9 & 0/9 \\
& Case~6 / 14B & 0/10 & N/A & N/A \\
\midrule
Unrequested $\Delta R(\ell,b)<0.5$ cut & Case~6 / 14B & 3/10 & 0/3 & 1/3 correct + 1/3 partial \\
& Case~6 / 32B & 0/10 & N/A & N/A \\
\bottomrule
\end{tabular*}
\caption{Implementation improvements from Case~2 to Case~6. ``Generation error'' denotes runs containing the corresponding implementation error, while ``Tracer exposure'' and ``Critique detection'' indicate how often the error was exposed or identified downstream.}
\label{tab:case2_case6_improvement}
\end{table}

\begin{table}[!ht]
\centering
\begin{tabular*}{0.95\textwidth}{@{\extracolsep{\fill}}p{0.30\textwidth}lccc@{}}
\toprule
Issue & Setting & Generation error & Tracer exposure & Critique detection \\
\midrule
Incorrect $\Delta\phi(\tau\tau,jj)$ & Case~5 / 14B & 10/10 & 10/10 & 10/10 \\
& Case~7 / 14B & 1/10 & 1/1 & 0/1 \\
\midrule
$b$-jet pre-filter instead of veto & Case~7 / 32B & 2/10 & 2/2 & 0/2 \\
\bottomrule
\end{tabular*}
\caption{Comparison of implementation errors and their detection for Cases~5 and 7. ``Generation error'' denotes runs containing the corresponding implementation error, while ``Tracer exposure'' and ``Critique detection'' indicate whether the error was exposed or identified downstream.}
\label{tab:case5_case7_comparison}
\end{table}

\subsection{Comparison of Golden Axe Oracle variants}
Finally, we evaluate the Golden Axe oracle performance. 
In this subsection, we describe the Golden Axe oracle variants,  \texttt{oracle-lite} and \texttt{oracle-lite-extensive}. The oracle prompts receive a task card and a helper registry. The task of the oracle is to identify important unresolved choices that may require clarification from the users. The \texttt{oracle-lite-extensive} is instructed to report every candidate and does not limit the number of claims. Some domain-independent instructions are given to the prompt to direct the model's attention to the nature of the collection scope. The outputs are the original phrases, a short description of the nature of the unresolved choice, and the reason why this may matter.  The \texttt{oracle-lite} shares almost the same prompts but is instructed to discard irrelevant candidates, such as when the choice is already resolved in the task card. It also merges candidates with the same underlying issue or minor issues. The number of claims is limited to at most three important choices. The output formats include a question to the user, two alternatives, and recommended defaults. Temperature is set to 0.2.

\begin{table}[!ht]
\centering
\begin{tabular*}{0.85\textwidth}{@{\extracolsep{\fill}}lccc@{}}
\toprule
Case & Runs & Total candidates & Candidates/run \\
\midrule
1 & 10 & 92 & 9.2 \\
2 & 10 & 83 & 8.3 \\
3 & 10 & 70 & 7.0 \\
4 & 10 & 82 & 8.2 \\
5 & 10 & 84 & 8.4 \\
6 & 10 & 89 & 8.9 \\
7 & 10 & 84 & 8.4 \\
\midrule
Total & 70 & 584 & 8.34 \\
\bottomrule
\end{tabular*}
\caption{Candidate counts from the exhaustive Oracle variant (\texttt{oracle-lite-extensive}) using Qwen3-32B over ten runs per case. Unlike the selective Oracle, no fixed upper limit is imposed on the number of reported candidates. }
\label{tab:oracle_v4_candidate_counts}
\end{table}

\begin{table}[!ht]
\centering
\begin{tabular*}{0.95\textwidth}{@{\extracolsep{\fill}}lccccc@{}}
\toprule
& \multicolumn{4}{c}{Knowledge beyond the helper registry} & \\
\cmidrule(lr){2-5}
Issue class & None & Low & Medium & High & Total \\
\midrule
Genuine unresolved tasks & 40 & 4 & 0 & 3 & 47 \\
Physics caution & 0 & 2 & 6 & 3 & 11 \\
Resolved by prompt & 110 & 9 & 14 & 10 & 143 \\
Invented alternative & 4 & 0 & 0 & 3 & 7 \\
\midrule
Total & 154 & 15 & 20 & 19 & 208 \\
\bottomrule
\end{tabular*}
\caption{Classification of the 208 issues reported by the prioritized Oracle (\texttt{oracle-lite}) using Qwen3-32B. Rows indicate the issue type, while columns denote the estimated amount of domain knowledge required beyond the user input, Oracle instructions, and helper registry. }
\label{tab:oracle_v5_issue_classification}
\end{table}

\begin{table}[!ht]
\centering

\begin{tabular*}{0.98\textwidth}{@{\extracolsep{\fill}}cp{0.52\textwidth}cc@{}}
\toprule
Case & Selected benchmark issue & Extensive  & oracle-lite  \\
& & detection & appearance \\
\midrule
1 & $b$-jet veto: selected vs.\ event-level jets & 9/10 & \textbf{1/29} \\
1 & Leading jets: selected vs.\ full jet collection & 10/10 & \textbf{10/29} \\
1 & Definition of the two-jet system & 9/10 & \textbf{9/29} \\
\midrule
2 & Hadronic $W$: selected vs.\ full non-$b$ jets & 10/10 & \textbf{10/30} \\
2 & Hadronic top: selected vs.\ full $b$ jets & 8/10 & \textbf{8/30} \\
2 & $\geq 2$ $b$-jets: collection scope & 5/10 & \textbf{0/30} \\
2 & Closest $b$-jet: collection scope & 4/10 & \textbf{0/30} \\
2 & Jet collection entering $H_T$ & 6/10 & \textbf{0/30} \\
2 & Jet collection entering the jet multiplicity & 5/10 & \textbf{0/30} \\
\midrule
5 & Opposite-sign tau pair / pairing order & 7/10 & \textbf{5/30} \\
5 & $b$-jet veto: collection scope & 8/10 & \textbf{2/30} \\
5 & ``summed momentum'' vs.\ four-momentum & 5/10 & \textbf{1/30} \\
5 & Definition of ditau centrality & 1/10 & \textbf{0/30} \\
\midrule
6 & Jet collection entering $H_T$ & 9/10 & \textbf{0/29} \\
6 & Jet collection entering the jet multiplicity & 6/10 & \textbf{0/29} \\
\midrule
7 & Opposite-sign tau pair / pairing order & 8/10 & \textbf{10/30} \\
7 & Interpretation of ``summed momentum'' & 9/10 & \textbf{0/30} \\
7 & Definition of ditau centrality & 1/10 & \textbf{1/30} \\
\bottomrule
\end{tabular*}
\caption{Comparison of issue recovery by the exhaustive Oracle variant (\texttt{oracle-lite-extensive}) and issue prioritization by \texttt{oracle-lite}, both using Qwen3-32B. Extensive-detection entries denote successful detections over ten runs, while the \texttt{oracle-lite} column gives matched issues relative to the total number of reported issues for each case. Cases~1 and 6 contain 29 reported issues because one run produced only two.}
\label{tab:oracle_v4_v5_issue_comparison}
\end{table}

We first show the results of the \texttt{oracle-lite-extensive} in Table~\ref{tab:oracle_v4_candidate_counts}.
The purpose of this comparison is to evaluate the coverage of predefined benchmark issues rather than the precision of the full candidate list. For oracle-lite-extensive, we count the number of runs in which each benchmark issue was detected. For oracle-lite, we count the reported candidates matching each benchmark issue, regardless of their classification in Table \ref{tab:oracle_v5_issue_classification}.
We set the target unresolved choice of each task card and record the number of runs detecting the ambiguities with Qwen3-32B. The total number of claims is listed in  Table \ref{tab:oracle_v4_candidate_counts}. 

For the \texttt{oracle-lite}, the number of claims is limited to no more than three. By comparing the reported candidates with those from \texttt{oracle-lite-extensive},  we can compare the claims that the LLM considers important with human-selected benchmark issues in the card. Ten runs for each of seven cases yield at most 210 claims. We obtained 208 claims and classified them as (the issue class) x(knowledge base) in Table \ref{tab:oracle_v5_issue_classification}. 
Here, “genuine unresolved” denotes valid issues that correspond to truly unresolved choices in the user input. “Resolved by prompt” denotes issues raised by the model even though they are already resolved in the prompt. “Invented alternative” denotes alternatives that are not supported by the provided inputs (the task card and helper registry) and are not useful for clarification. “Physics caution” denotes valid and potentially useful concerns that go beyond the provided inputs. We also classify each claim according to the knowledge required to produce it. The provided knowledge base consists of the user input (task card), the Oracle instructions, and the helper registry. “None” indicates that the claim can be derived directly from these inputs, for example by checking object scope, processing order, or consistency of named objects. “Low” indicates that only a small amount of conventional domain knowledge is needed beyond the provided knowledge base. “Medium” denotes claims that rely on a recognizable analysis convention or physics prior not explicitly supplied in the inputs, while “High” denotes claims that introduce more specific reconstruction choices, experimental conventions, or physics assumptions that are not supported by the provided knowledge base.
Of the 208 claims reported by oracle-lite, 47 concerned genuinely unresolved choices in the task cards, 11 were physics cautions requiring additional domain knowledge, 143 concerned choices already resolved in the cards, and 7 introduced unsupported alternatives. Thus, the dominant limitation of this configuration is the repeated reporting of already specified choices, rather than the invention of entirely unsupported alternatives.

Table \ref{tab:oracle_v4_v5_issue_comparison} shows the coverage of the selected benchmark issues by \texttt{oracle-lite}. The \texttt{oracle-lite} claims overlap partly with the selected benchmark issues but remain independently generated.
\bibliographystyle{JHEP}
\bibliography{biblo}

\end{document}